\documentclass[onecolumn]{aastex631}

\usepackage{graphicx} 
\usepackage{rotating}
\usepackage[caption = false]{subfig}
\graphicspath{{figures/}}

\usepackage[]{hyperref}
\hypersetup{colorlinks = true, linkcolor = blue, anchorcolor = red, citecolor = blue, filecolor = red, urlcolor = red, hypertexnames=true}

\received{\today}
\revised{\today}
\accepted{\today}
\submitjournal{JAAVSO}
\shorttitle{Exoplanet WASP-140b}
\shortauthors{North \& Banks}

\begin{document}

\title{Photometry and transit modeling of exoplanet WASP-140b}
\correspondingauthor{Timothy Banks}
\email{tim.banks@nielsen.com}

\author{Allen North}
\affiliation{Department of Physical Science \& Engineering, Harper College, 1200 W Algonquin Rd, Palatine, IL 60067, USA}

\author{Timothy Banks}
\affiliation{Department of Physical Science \& Engineering, Harper College, 1200 W Algonquin Rd, Palatine, IL 60067, USA}
\affiliation{Data Science, Nielsen, 200 W Jackson, Chicago, IL 60606, USA}


\begin{abstract}
\noindent
Eleven transit light curves for the exoplanet WASP-140b were studied with the primary objective to investigate the possibility of transit timing variations (TTVs).  Previously unstudied MicroObservatory and Las Cumbres Global Telescope Network photometry were analysed using Markov Chain Monte Carlo techniques, including new observations collected by this study of a transit in December 2021. No evidence was found for TTVs.  We used two transit models coupled with Bayesian optimization to explore the physical parameters of the system. The radius for WASP-140b was estimated to be $1.38^{+0.18}_{-0.17}$ Jupiter radii, with the planet orbiting its host star in $2.235987 \pm 0.000008$ days at an inclination of $85.75 \pm 0.75$ degrees. The derived parameters are in formal agreement with those in the exoplanet discovery paper of 2016, and somewhat larger than a recent independent study based on photometry by the TESS space telescope.

\end{abstract}


\keywords{Exoplanets --- Transits}

\section{Introduction} \label{sec:intro}

An exoplanet is, in general, a planet orbiting a star other than our Sun. The first confirmed discoveries of exoplanets were made in the early 1990s, opening up a field that is rapidly expanding with several thousand confirmed exoplanets known today, giving us insight into different planetary systems to our own and introducing challenges to our understanding of how such systems form and evolve. 

A variety of techniques are used to discover exoplanets. In this project, we concentrated on the transit methods that have been used to discover the most exoplanets to date --- namely monitoring the brightness of the exoplanet system. Exoplanets are generally too close to their host stars to be seen as separate objects. The transit method tracks the brightness of the combined system (exoplanets and host star) with time, looking for changes caused such as when the planet passes in front of its star and blocks some light from reaching the Earth. The method tells us about the size of the planets and the angle they orbit about the host star relative to our line of sight.

In this paper we study transits for the exoplanet WASP-140b.  This planet was discovered by Hellier {\em et al.} (2016), being 2.4 Jupiter masses orbiting its V=11.1 K0 host star (coordinates $\alpha = 04^{h} 01^{m} 32.54^{s}$, $\delta = -20^{\circ}27' 03.9"$ J2000) once in roughly 2.24 days.  Hellier {\em et al.} note a rotational modulation of the out of transit flux with an $\sim10.4$ day cycle, which they attribute to magnetic activity of the host.  They note that the transit is grazing, leading to a higher uncertainty in the estimate radius of the planet ($1.44^{+0.42}_{-0.18}$ Jupiter radii). 

We apply the {\sc exotic} model (Zellem {\em et al.}, 2020) to estimate basic parameters of the system such as time of mid-transit, planetary radius relative to the host star, and orbital radius.  We compare and contrast these results with a simple transit model (Mandel \& Agol, 2002) we implemented with a Bayesian optimizer, as well as with literature results. We were particularly interested in seeing if there were deviations in the times of mid-transits compared to a fixed orbital period.  The Transit Timing Variation (TTV) method is based on monitoring such changes in timing of transits. The presence of non-transiting planets (in the same system) can be inferred from TTV measurements. The gravitational interaction of these non-transiting planets will sometimes increase the orbital period of the transiting planet, and at other times decrease the period, depending on their relative postions and so the mid-transit times will vary from a fixed, regular cycle.


%
\begin{figure}[htb]
    \centering 
\begin{subfloat}[18 November 2018]{
  \includegraphics[width=0.45\textwidth]{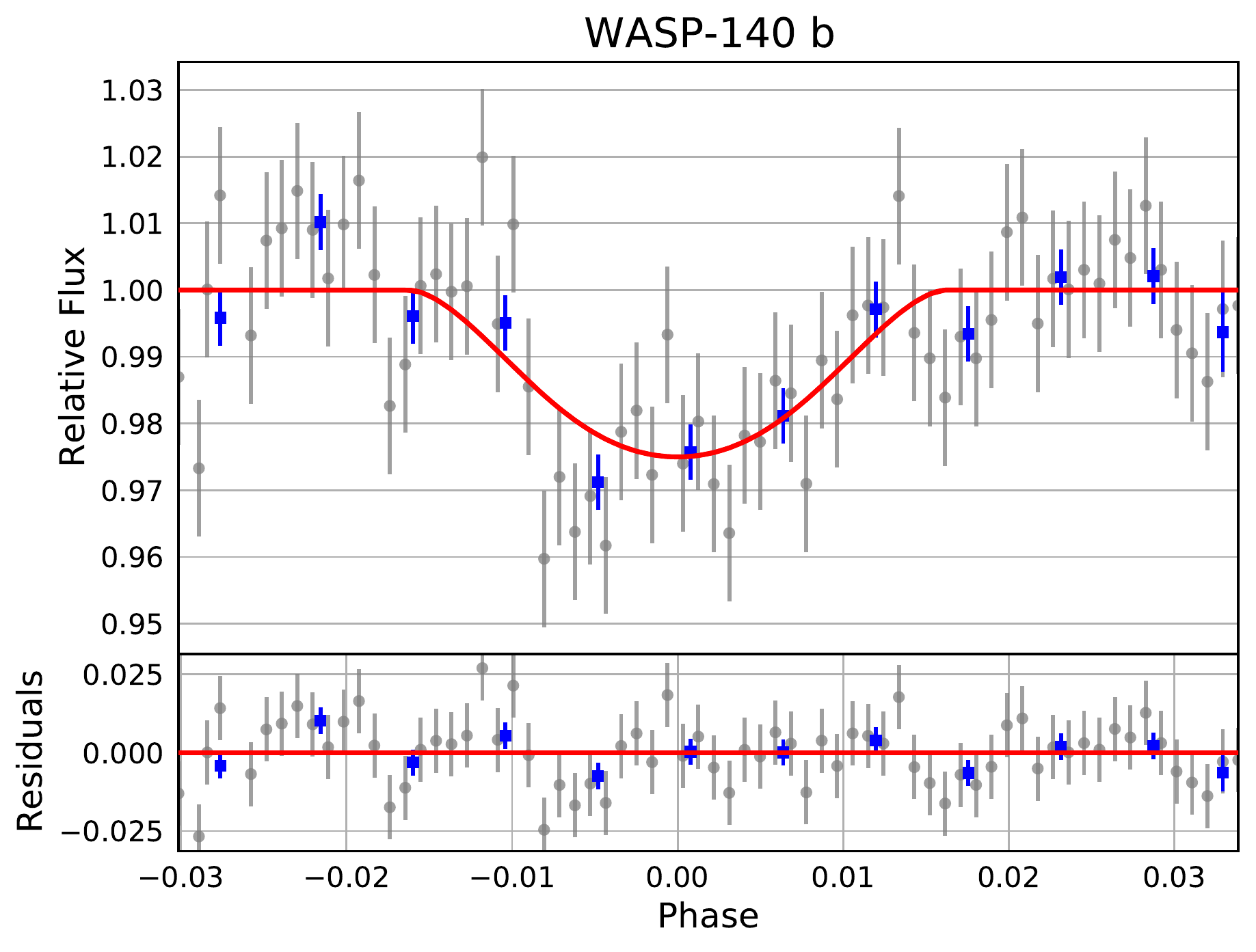}
  \label{fig:1}}
\end{subfloat}\hfil 
\begin{subfloat}[22 January 2019]{
  \includegraphics[width=0.45\textwidth]{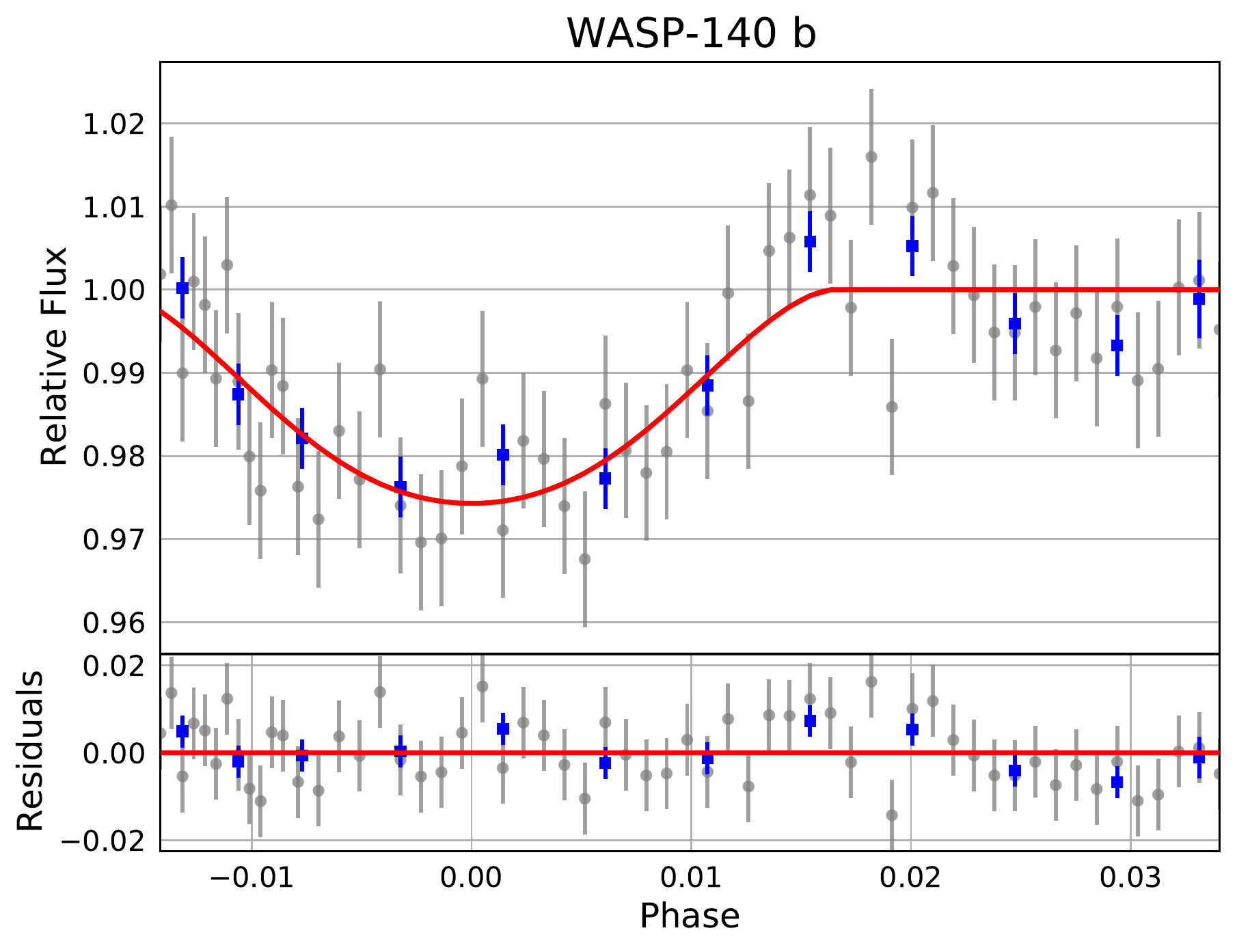}
  \label{fig:2}}
\end{subfloat}\hfil 
\medskip

\begin{subfloat}[11 October 2020]{
  \includegraphics[width=0.45\textwidth]{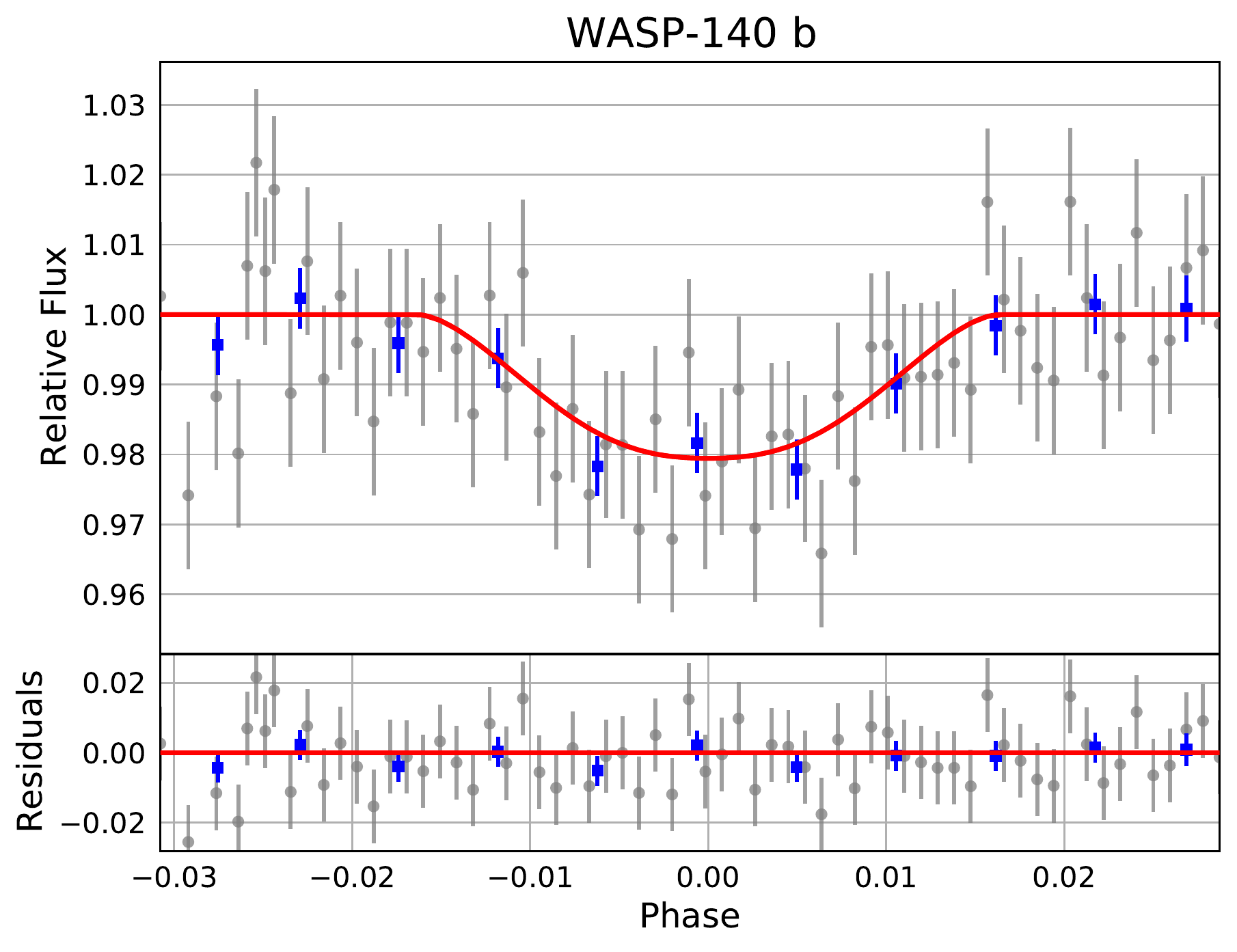}
  \label{fig:3}}
\end{subfloat}
\begin{subfloat}[20 October 2020]{
  \includegraphics[width=0.45\textwidth]{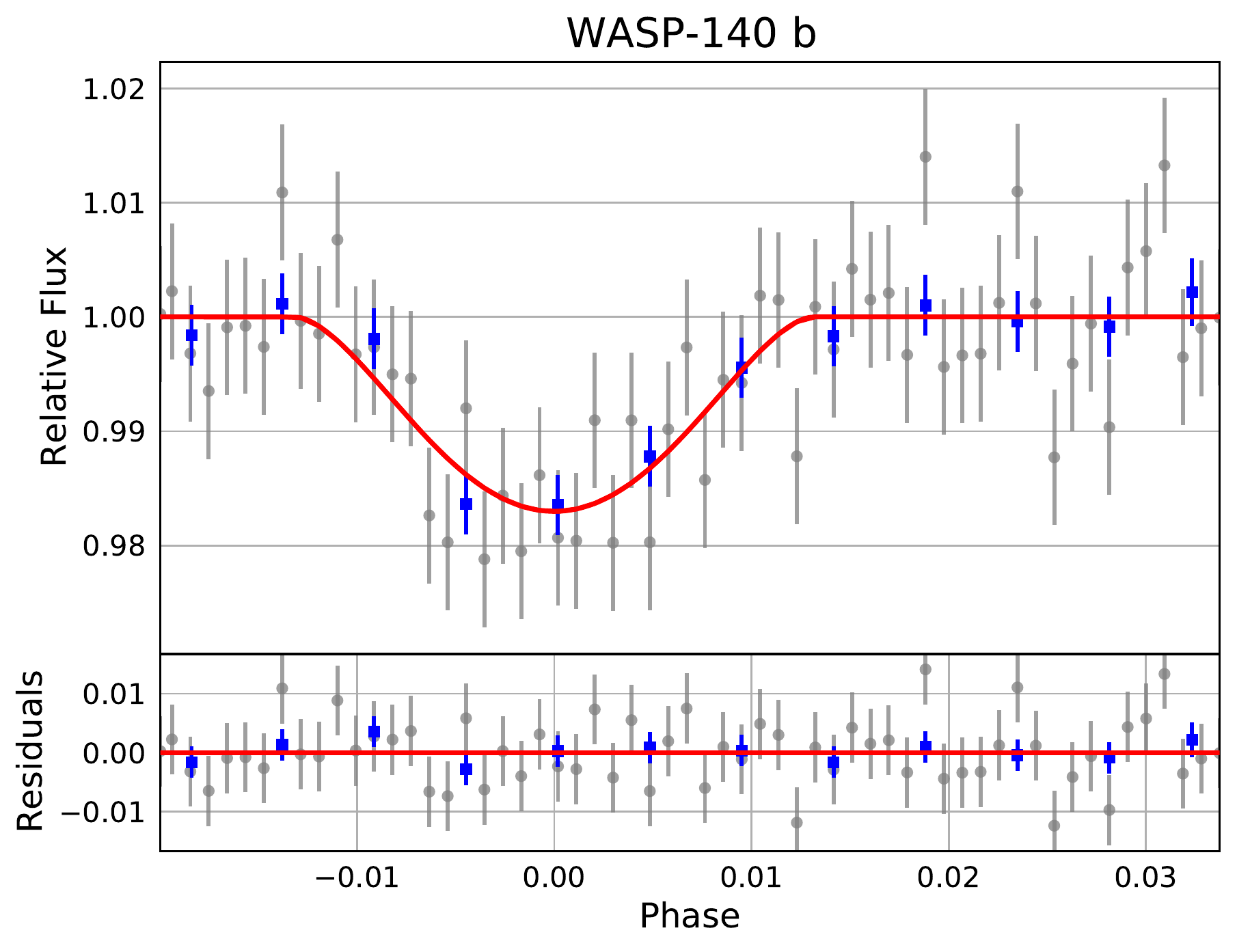}
  \label{fig:4}}
\end{subfloat}\hfil 
\medskip

\begin{subfloat}[29 October 2020]{
  \includegraphics[width=0.45\textwidth]{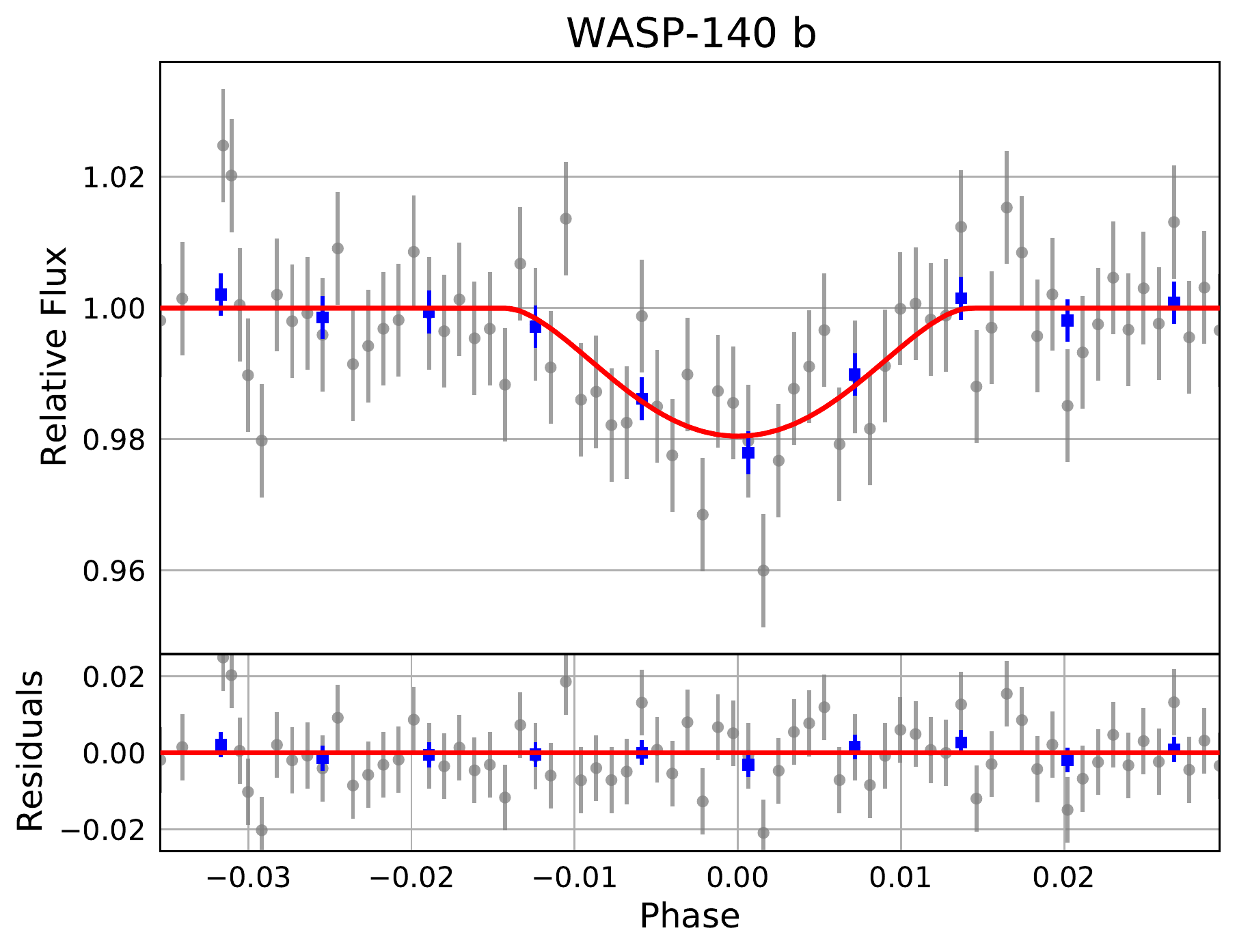}
  \label{fig:5}}
\end{subfloat}\hfil 
\begin{subfloat}[02 January 2021]{
  \includegraphics[width=0.45\textwidth, height=0.34\linewidth]{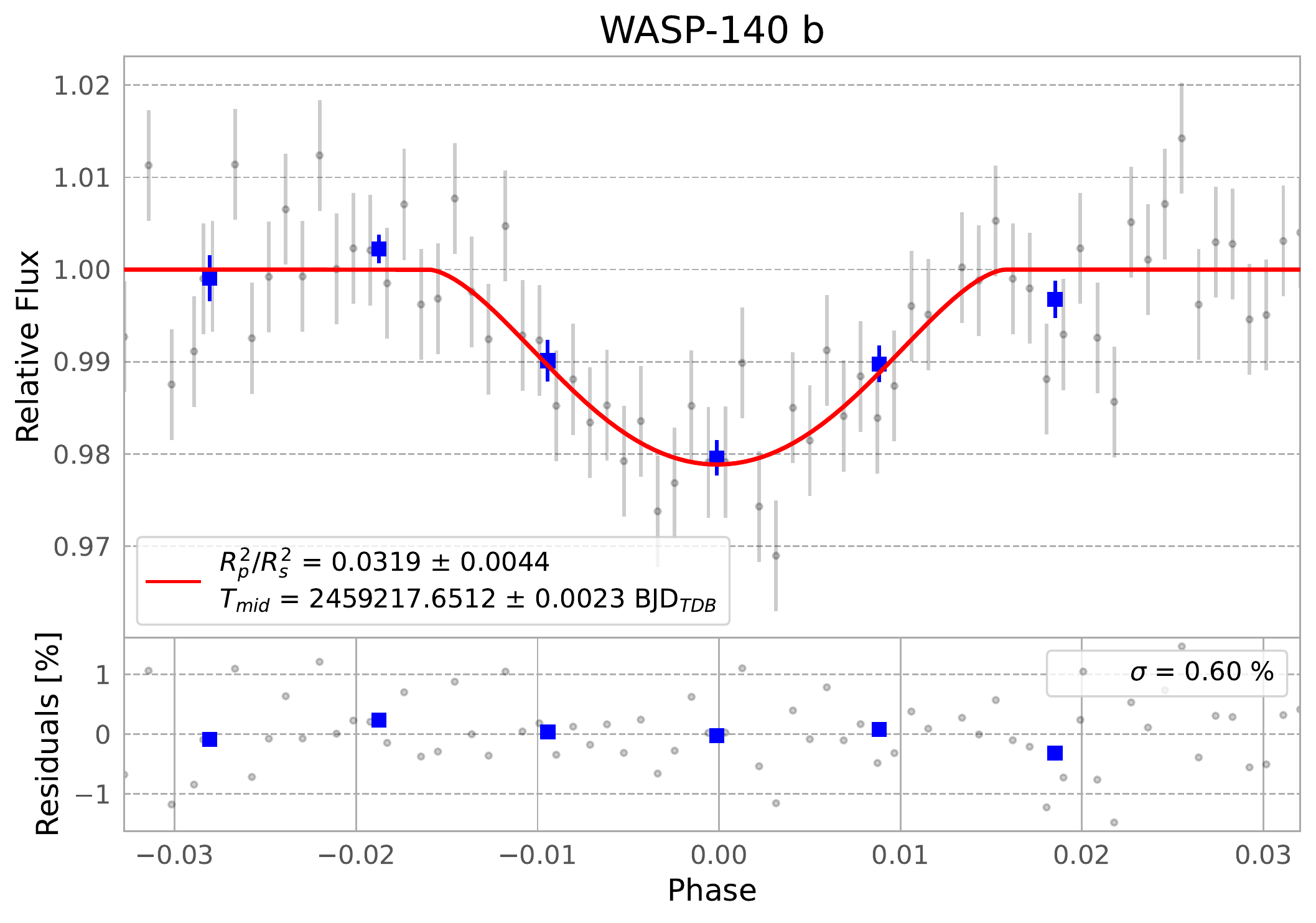}
  \label{fig:7}}
\end{subfloat}\hfil 

\caption{Selected WASP-140b transit data collected by the MicroObservatory and models.  MicroObservatory observations have no filter. The red lines show the expected variation based on the best fitting {\sc exotic} model for each transit. Not all transits are shown for reasons of space.
\label{fig:MObs_WASP_140_transits}}
\end{figure}

\begin{sidewaysfigure}[htb]
\centering 
\begin{subfloat}[04 October 2019 ($w$)]{
  \includegraphics[width=0.45\linewidth]{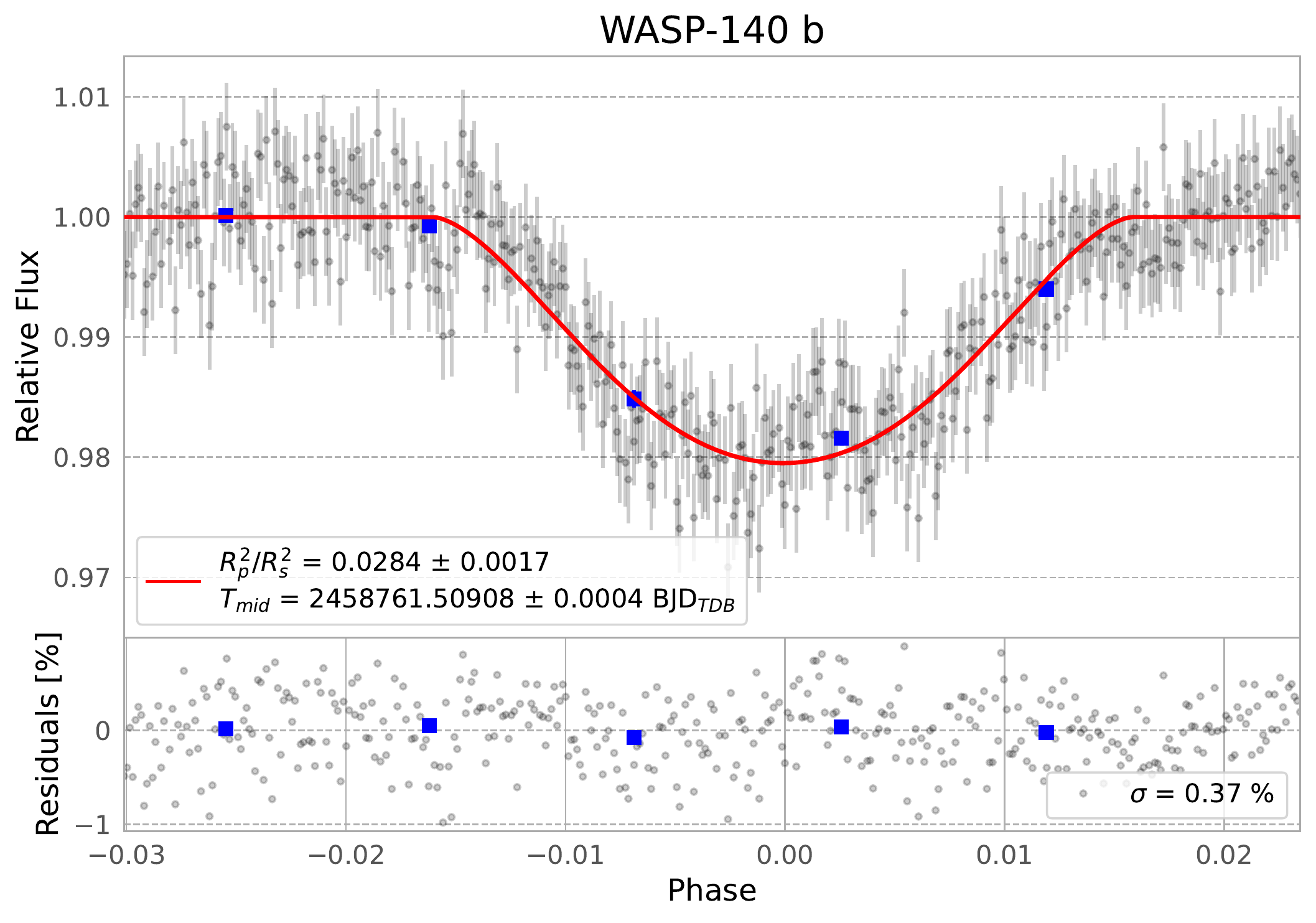}
  \label{fig:8}}
\end{subfloat}\hfil 
\begin{subfloat}[14 October 2020 ($i_p$)]{
  \includegraphics[width=0.45\linewidth]{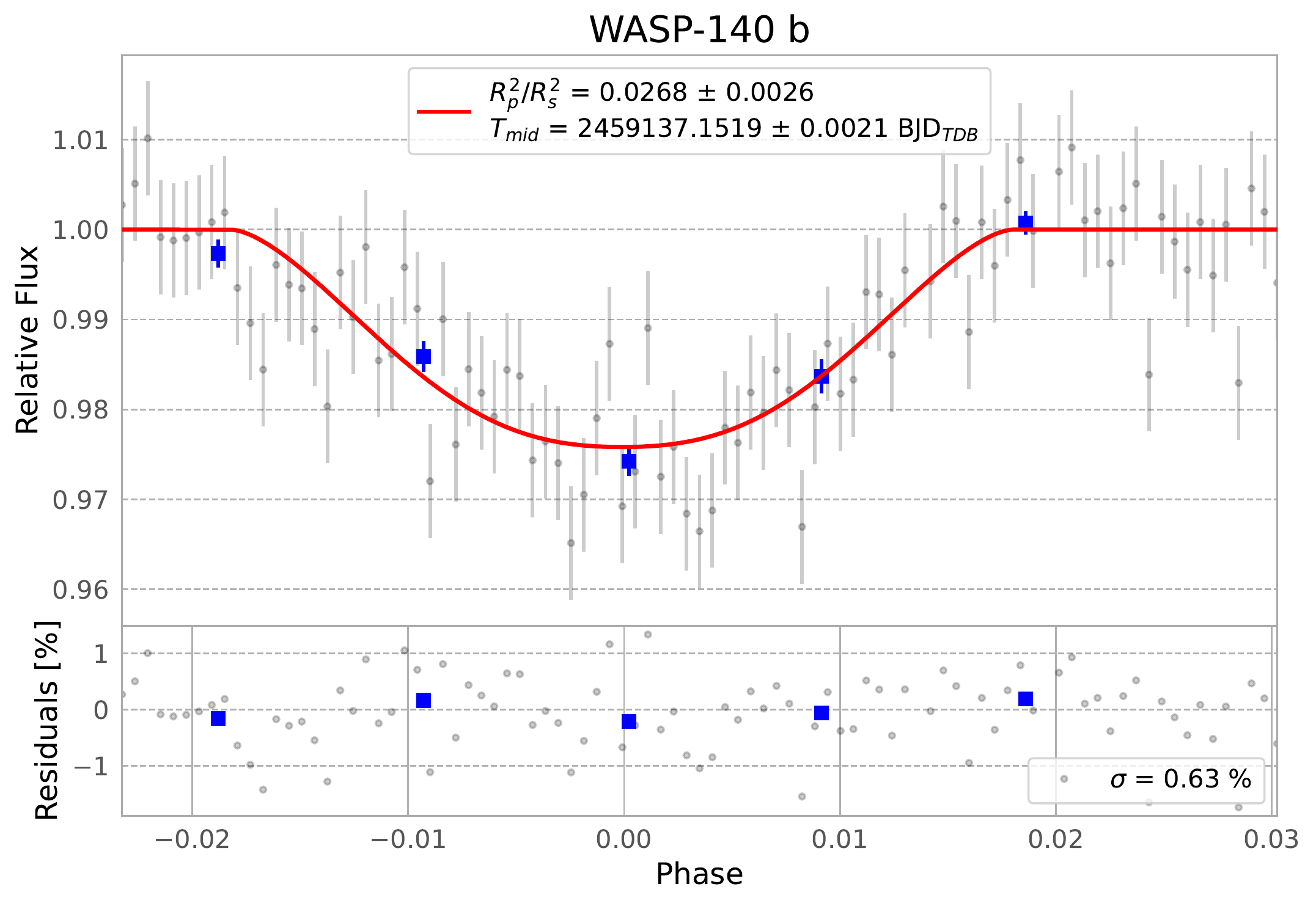}
  \label{fig:9}}
\end{subfloat} 
\medskip

\begin{subfloat}[24 October 2021 ($i_p$)]{
  \includegraphics[width=0.45\linewidth]{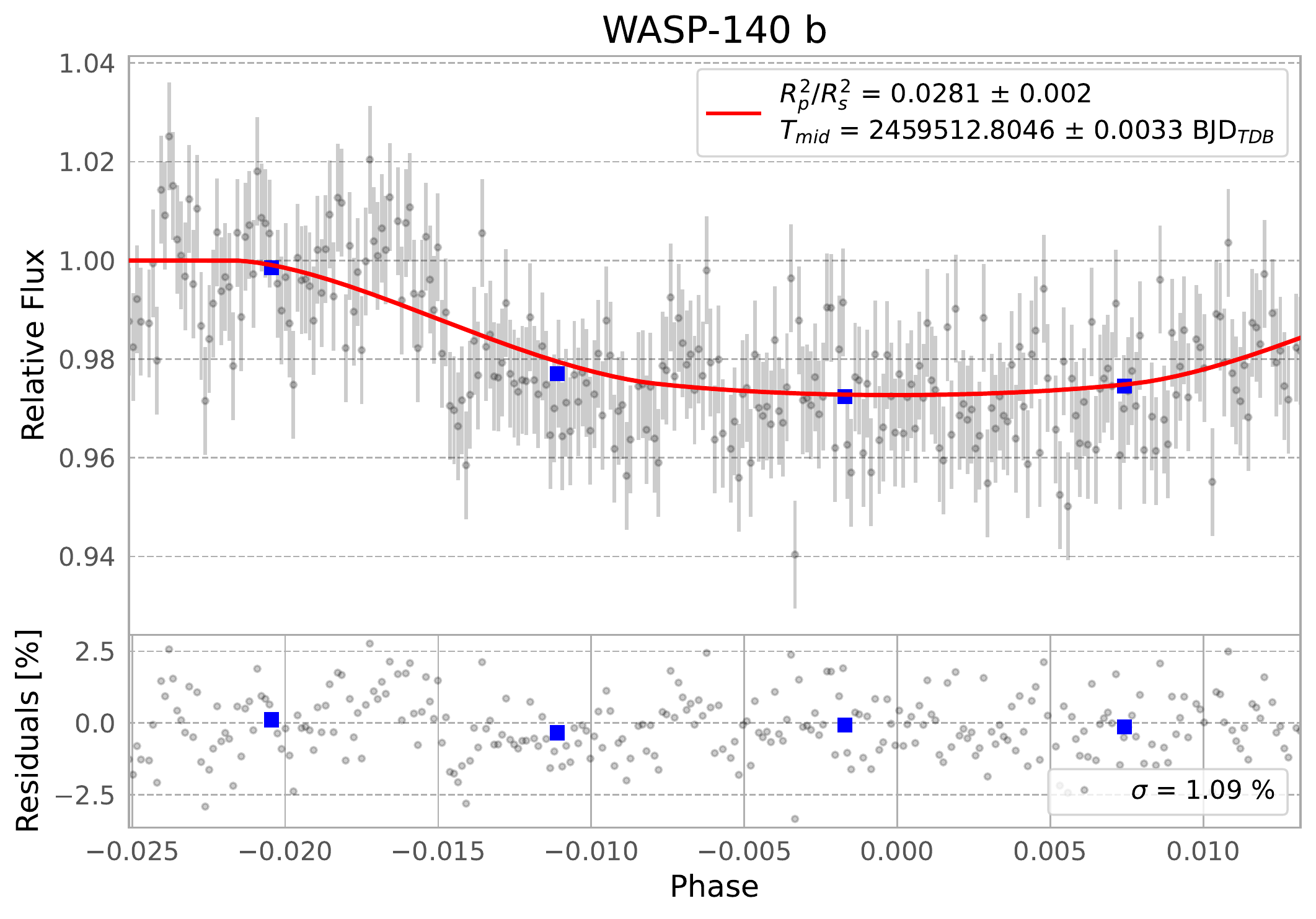}
  \label{fig:10}}
\end{subfloat}\hfil 
\begin{subfloat}[28 December 2021 ($r_p$)]{
  \includegraphics[width=0.45\linewidth]{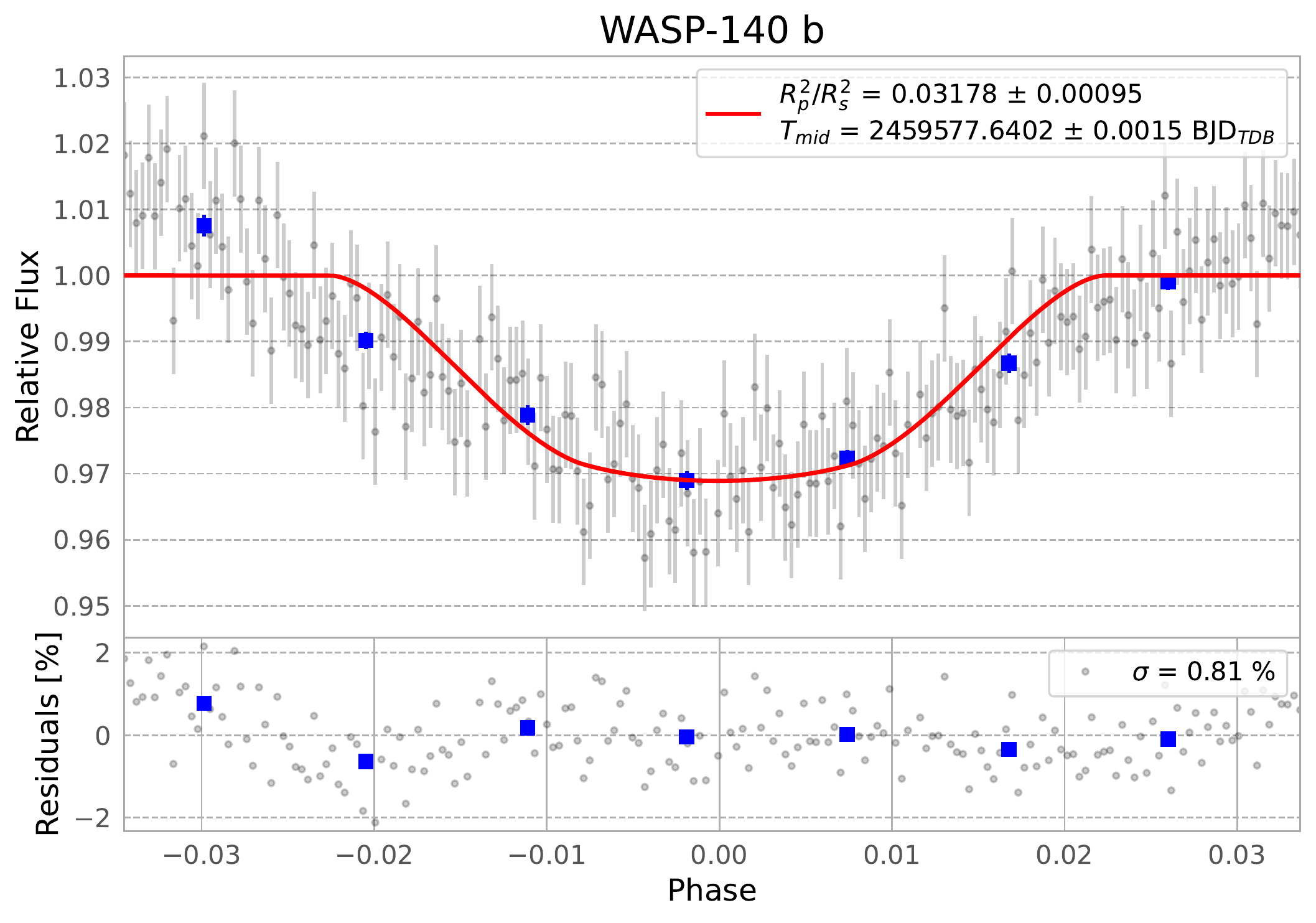}
  \label{fig:11}}
\end{subfloat}
\caption{WASP-140b transit data collected using the LCO. The filters used for the LCO observations are indicated in the appropriate sub-figure captions. 
\label{fig:LCO_WASP_140_transits}}
\end{sidewaysfigure}

\begin{table}[t]
    \caption{{\bf Fitted Parameters for WASP-140b} from the EXOTIC modelling. Mid-transit times are given in Barycentric Julian Dates (Barycentric Dynamical Time), the orbital semi-major axis ($a$) in terms of the stellar radius ($r_s$), and the planetary radius ($r_p$) relative to the stellar radius.  {\sc exotic} outputs ${a}/{r_{s}}$, so a column giving the inverse is given for convenience when comparing with a later model and the literature.  Uncertainties are $1\sigma$. `Quality' is a subjective assessment by the authors of the quality of the light curve. Exposure times for the LCOGT observations were 16.5 seconds for 4 October 2019, 100 s for 14 October 2020, 16.8 s for 24 October 2021, and 60 seconds for 28 December 2021.}
    \centering
    \hspace{-2.5cm}
    \begin{tabular}{||l|l|l|l|l|l||}
    \hline
    Date         &      Mid-transit            & ${a}/{r_{s}}$         & $r_{s}/a$                  & $r_{p} / r_{s}$      & Quality  \\
    \hline
    18 Nov 2018  &  2458441.7633  $\pm$ 0.0028 & 7.69  $\pm$ 0.30      & $0.130 \pm 0.005$          & 0.1786  $\pm$ 0.0099 & complete \\
    22 Jan 2019  &  2458506.6080  $\pm$ 0.0026 & 7.63  $\pm$ 0.24      & $0.131 \pm 0.004$          & 0.179   $\pm$ 0.001  & partial  \\
    11 Oct 2020  &  2459134.9220  $\pm$ 0.0038 & 7.51  $\pm$ 0.52      & $0.133^{+0.010}_{-0.009}$  & 0.154   $\pm$ 0.024  & complete \\
    20 Oct 2020  &  2459143.8611  $\pm$ 0.0020 & 8.40  $\pm$ 0.26      & $0.119 \pm 0.004$          & 0.178   $\pm$ 0.015  & complete \\
    29 Oct 2020  &  2459152.8145  $\pm$ 0.0026 & 8.14  $\pm$ 0.33      & $0.123 \pm 0.005$          & 0.176   $\pm$ 0.016  & complete \\
    15 Dec 2020  &  2459199.7704  $\pm$ 0.0083 & 7.29  $\pm$ 0.64      & $0.137^{+0.013}_{-0.011}$  & 0.119   $\pm$ 0.030  & partial  \\
    02 Jan 2021  &  2459217.6512  $\pm$ 0.0023 & 7.70  $\pm$ 0.24      & $0.130 \pm 0.004$          & 0.179   $\pm$ 0.012  & complete \\
    04 Oct 2019  &  2458761.5091  $\pm$ 0.0004 & 7.631 $\pm$ 0.085     & $0.131 \pm 0.001$          & 0.1684  $\pm$ 0.005  & complete \\
    14 Oct 2020  &  2459137.1516  $\pm$ 0.0023 & 7.20  $\pm$ 0.23      & $0.139^{+0.005}_{-0.004}$  & 0.1618  $\pm$ 0.0085 & complete \\
    24 Oct 2021  &  2459512.8046  $\pm$ 0.0033 & 6.56  $\pm$ 0.19      & $0.152^{+0.005}_{-0.004}$  & 0.1678  $\pm$ 0.0059 & partial  \\
    28 Dec 2021  &  2459577.6402  $\pm$ 0.0015 & 6.486 $\pm$ 0.035     & $0.154 \pm 0.001$          & 0.1783  $\pm$ 0.0027 & partial  \\
    \hline 
    \end{tabular} 
    \label{tab:exotic_wasp_140b_fits}
\end{table}

\section{Data and Initial Processing}

The bulk of observations are 60-second, unfiltered exposures collected by a 6-inch aperture MicroObservatory (MObs; Sadler {\em et al.}, 2001) telescope located at Mount Hopkins (latitude $31.675^\circ$, longitude $-110.952^\circ$, 1,268m altitude above sea level) in Arizona, using a KAF-1403 ME CCD camera with a pixel scale of 5.2" per pixel and $2 \times 2$ binning to reduce noise.  These data were analysed using {\sc exotic}, which is a {\sc python}-based tool developed by JPL's `Exowatch' program for reducing exoplanet transit data. This software can run on a variety of operating systems as well as via Google's online `Colaboratory'\footnote{For further details on this tool see: https://research.google.com/colaboratory/faq.html} tool. Technical details on {\sc exotic} can be found in Zellem~{\em et al.} (2020). Priors for Markov Chain Monte Carlo (MCMC) fitting by {\sc exotic} are automatically scraped from the NASA Exoplanet Archive (Akeson~{\em et al.}, 2013), while limb darkening parameters are generated by {\sc exofast} (Eastman~{\em et al.}, 2013). {\sc exotic} generates $1\sigma$ uncertainties based on the resulting posterior distributions. 

Only dark images were available for the MObs observations, i.e., no flat field images were collected. The dark frames were collected at the beginning and end of each night of observation. As part of the analysis, {\sc exotic} applied the dark frames to the science data, and then performed differential aperture photometry.  For each transit, the analyst supplied {\sc exotic} a list of comparison stars. {\sc exotic} performed a stability assessment of this candidate list, choosing the most stable star as the final comparison star.  Relatively poor pointing accuracy of the telescope and drift in tracking throughout a transit could lead to selection of different final comparison stars across the transits. However, typically {\sc exotic} selected stars 108 or 112 from the AAVSO comparison star sequence for WASP-140. We plate-solved science frames for each transit to ensure correct selection of the exoplanet host star, using astrometry.net, together with confirmation using charts prepared using the online AAVSO finding chart tool.


\section{Analysis}

We analysed 22 MObs attempts to observe transits of WASP-140b, dating from 12 October 2016 to 24 October 2021. Only 7 resulted in successful measurements of transits (see Figure~\ref{fig:MObs_WASP_140_transits} for charts of representative transits), a success rate of 32\%. Clouds or incorrect pointing of the telescope accounted for the failed attempts. Table~\ref{tab:exotic_wasp_140b_fits} lists the key output from these fits using {\sc exotic}, namely the orbital semi-major axis $a$ (relative to the stellar radius $r_s$), the planetary radius ($r_p$), and the time of mid-transit (in BJD).  The observations and fitted parameter values from {\sc exotic} have been uploaded to the AAVSO exoplanet database, under the usercode BTSB.

We also made use of the Las Cumbres Observatory Global Telescope network (LCOGT; Brown~{\em et al.}, 2013), first using archival data of transits and also collecting $r_p$ photometry on the night of 28 December 2021 using a telescope at the Cerro Tololo Inter-American Observatory. All the analysed LCOGT were collected using 0.4 meter telescopes equipped with CCDs. We processed all these data using {\sc exotic}, following flat fielding, dark subtraction, and bias correction via the LCO {\sc banzai} system.\footnote{See https://github.com/LCOGT/banzai for further information on this data pipeline.}  Model fits to the transits are shown in Figure~\ref{fig:LCO_WASP_140_transits} and final parameter estimates are given in Table~\ref{tab:exotic_wasp_140b_fits}.  We did not upload the LCOGT archival data or the model fits based on these to the AAVSO exoplanet database, given that we did not collect the data and did not wish to `make claim' to them over the original investigators.


\begin{figure}  
\centerline{\includegraphics[height=8.5cm]{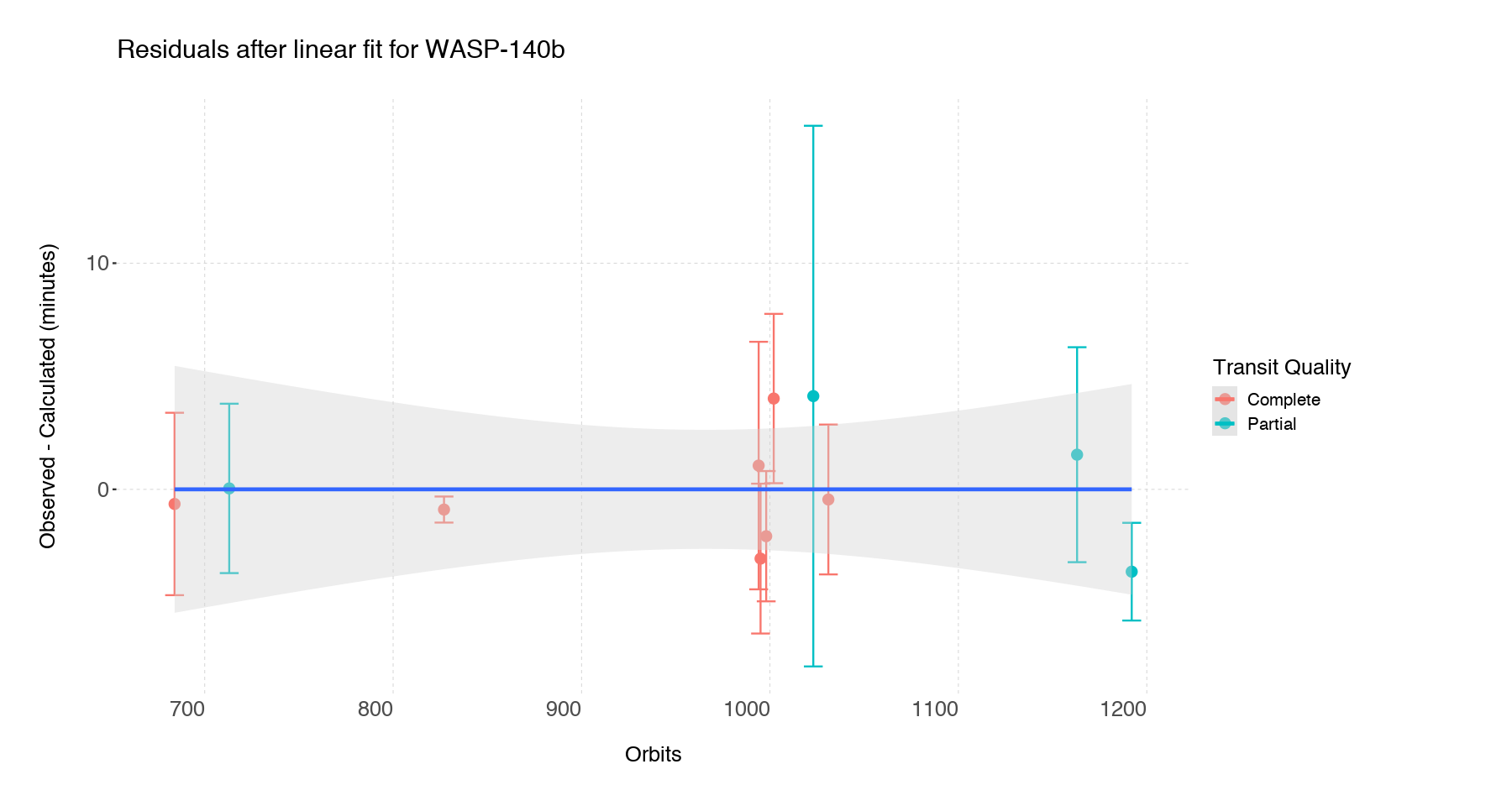}}
\caption{Residuals from linear regression fit of orbits versus mid-transit time for WASP-140b. A linear model was fitted to the residuals, with no statistically significant slope.  The grey shaded zone is the $3 \sigma$ confidence interval for the regression.  The blue line is the mean regression slope, which is not statistically different from zero at the $3 \sigma$ level. The error bars for the mid-transit timing estimates overlap with this, and with zero, indicating no statistically significant trends in the residuals.  Transits were classified by eye into complete and incomplete transits, to see if data quality might obscure any trends (see Table~\ref{tab:exotic_wasp_140b_fits}). It does not.
\label{fig:wasp_140b_residuals} 
}
\end{figure}

\subsection{Orbital Period}

The ephemeris of Hellier {\em et al.} (2016) was used to calculate the number of orbits made by Wasp-140b about its host star since their starting epoch.  These were then regressed against the mid-transit times given in Table~\ref{tab:exotic_wasp_140b_fits} using the `lm' function in R (R Core Team, 2021),\footnote{R is available from https://www.r-project.org} giving an orbital period of $2.235987 \pm 0.000008$ days and an epoch of $2456912.349 \pm 0.008$.  These are in good agreement with the values of Hellier {\em et al.} (2016): $2.2359835  \pm 0.0000008$ days for the orbit and $2456912.35105 \pm 0.00015 $ for the epoch. Higher order polynomial fits did not result in additionally statistically significant parameters. Inspection of the residuals (see Figure~\ref{fig:wasp_140b_residuals}) reveals no apparent variation in period.  These results therefore do not indicate any significant transit timing variations (TTVs).  As noted above, TTVs would indicate the presence of an additional planet in the WASP-104 system through its gravitational attraction periodically altering the orbital velocity of WASP-140b.  This would have led to observed transits (of WASP-140b) being earlier or later than predicted by a linear ephemeris.  Maciejewski (2022) also analysed Transiting Exoplanet Survey Satellite (TESS, Ricker~{\em et al.}, 2015) data for the system searching unsuccessfully for TTVs, concluding that there were none currently detectable and so in agreement with the current study.


\begin{sidewaysfigure}  
\centerline{\includegraphics[height=0.54\textheight]{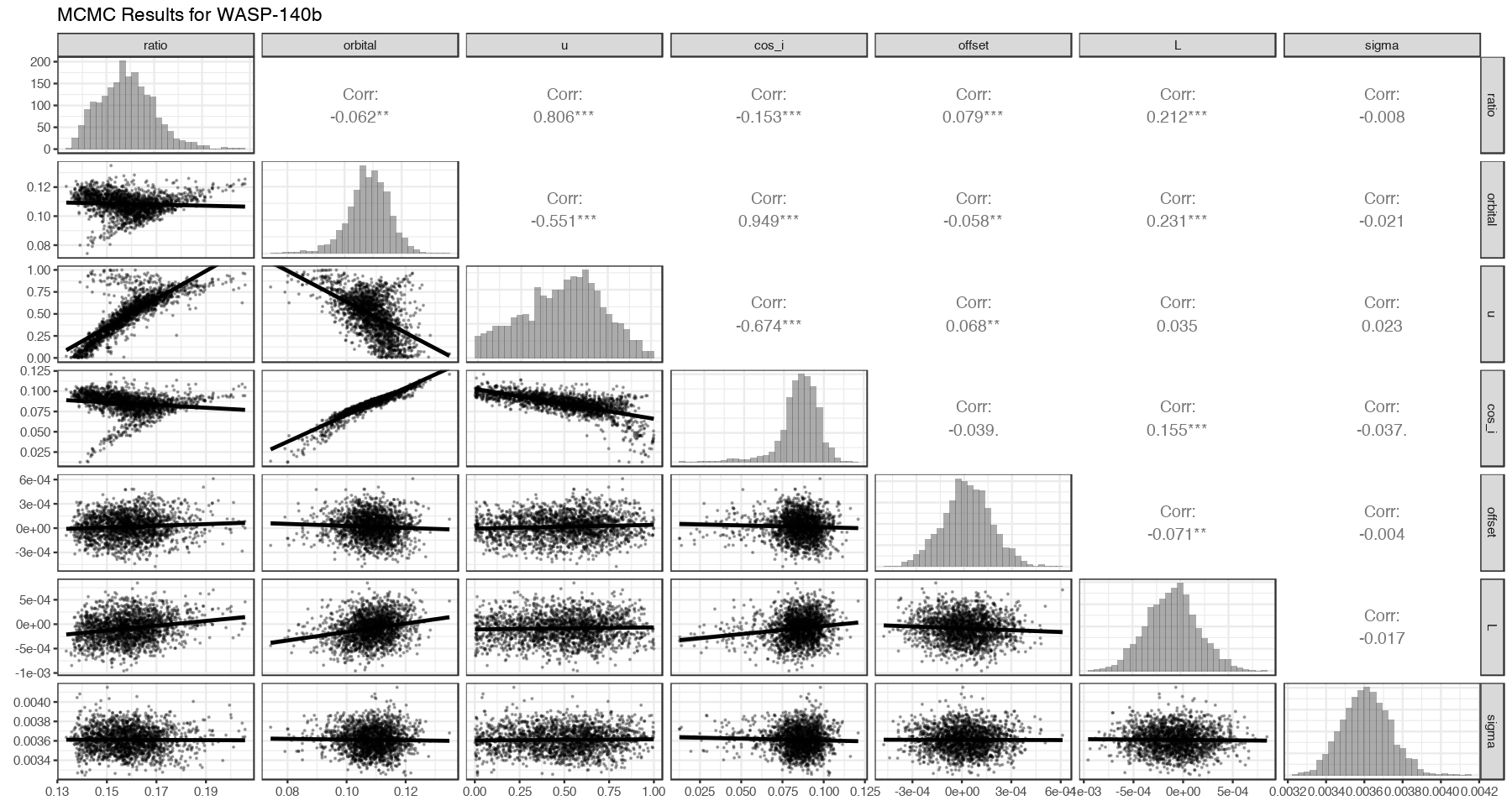}}
\caption{
Example MCMC results for the 4 October 2019 transit of WASP-140b. This represents 4,000 steps in the Markov chain, including the initial steps known as `burn-in'.  These steps are excluded from the final results, and are considered a result of starting the optimization in a lower probability set of parameters, leading to movement to the global minimum. Actual runs included 40,000 steps, which unfortunately `overloaded' the plotting software and are therefore not included here. `Ratio' is the ratio of the planetary radius to the stellar one, `orbital' is the ratio of stellar radius to the orbital semi-major axis, `u' is the linear limb darkening co-efficient, `cos\_i' is the cosine of the inclination, 'offset' an adjustment in phase, `L' an adjustment in flux, and `sigma' an estimate of the white noise in the data. The chart provides the distributions of each of these parameters on its diagonal as bar charts, correlations between the variables are given in the upper right, and scatter plots crossing each of the parameters in turn are given in the lower left.  Each point in a scatter plot represents a step in the Markov chain.  The bold lines are linear regressions to the data, corresponding to the correlation results. 
\label{fig:14_oct_19_wasp_140_mcmc}  
}
\end{sidewaysfigure}

\subsection{Transit Models}

While {\sc exotic} had already fitted the transits, we decided to build from `first principles' a simple transit model and couple this with optimization techniques in order to both make a comparison and explore including inclination as a free parameter.  This was primarily a student project acting as an introduction to exoplanet research, so building our own model and coupling this with optimization was considered a good learning exercise. {\sc exotic} adopts its priors from the NASA Exoplanet Archive, hence it adopted the inclination from Hellier {\em et al.} (2016) as a fixed parameter.  Given the glancing nature of this transit, fixing the inclination has a large effect on the derived parameter estimates. For optimization of our transit model, we used the Markov Chain Monte Carlo (MCMC) technique Hamilton Monte Carlo (HMC). MCMC allows construction of a Markov process such that the stationary distribution is the same as our target distribution,  through the generation of a `chain' of random samples from the process.  Through a sufficient number of samples, such a chain becomes close enough to the stationary distribution and therefore provides a good approximation to the target distribution. This is known as convergence of the MCMC chain (see Sinharay, 2003), and allows exploration of the uncertainty in the parameter estimates --- explaining our interest in this technique.  We implemented HMC using the {\em rstan}\footnote{Available from https://mc-stan.org/users/interfaces/rstan} implementation of Stan (Carpenter~{\em et al.}, 2017; Stan Development Team, 2016) inside the statistical programming language R. Uniform priors were used, reflecting minimum previous knowledge of the parameters.

To build this model we used some key parameters of the exoplanet and its host star:
\begin{itemize}
    \item{$a$, $r_s$, and $r_p$ were as defined above, with the radii being in terms of $a$;}
    \item{$u$ = linear limb darkening coefficient (see below for an explanation of this parameter);}
    \item{$i$ = orbital inclination (in degrees). Ninety degrees means that the orbital plane is in the line of sight from the Earth;}
    \item{{\em offset} = a parameter to adjust the reference point of phase axis;}
    \item{$U$ = system brightness, used to adjust the reference point of flux axis. The out of transit flux should be approximately unity, i.e., the fluxes are normalized to the mean out of transit level.}
\end{itemize}
We first consider that $d$ is the center-to-center distance between the planet and the star. We can then calculate $z = \frac{d}{r_*}$, which denotes the normalised separation of the centers (of the exoplanet and its host star) and $p = \frac{r_p}{r_*}$, which is the ratio of the disk radii. This allows us to model a transit based on the equations in Mandel \& Agol's (2002) paper. These specify that for a uniform source, the ratio of obscured to unobscured flux is $F^e(p, z)=1-\lambda^e(p, z)$, where
\begin{equation}\label{eqn:mandel}
    \lambda^e_{(p, z)}=\left\{\begin{array}{ll}
    0                            & 1+p<z \\
    \frac{1}{\pi}\left[p^{2} k_{0}+k_{1}-\sqrt{\frac{4 z^{2}-\left(1+z^{2}-p^{2}\right)^{2}}{4}}\right] 
    & |1-p|<z \leq 1+p \\
    p^{2}                        & z \leq 1-p \\
    1                            & z \leq p-1.
    \end{array}\right. 
\end{equation}
and $\kappa_{1}=\cos ^{-1}\left[\left(1-p^{2}+z^{2}\right) / 2 z\right]$ and $\kappa_{0}=\cos ^{-1}\left[\left(p^{2}+z^{2}-1\right) / 2 p z\right] .$ 
This set of equations describe the flux of planetary systems in the following cases:
\begin{enumerate}
    \item{When the planetary disk does not obscure any portion of the stellar disk. There will be no dimming of the combined light, and so the normalized flux would be 1.}
    \item{When the planetary disk is completely in front of the stellar disk. In the case of a uniformly bright stellar disk, the dimming will scale by the obscured area -- which can be calculated by $\frac{r_{p}^2}{r_{s}^2}$ (or $p^2$).}
    \item{The boundary case when the planetary disk is moving onto or off the stellar disk.}
\end{enumerate}
The fourth case in Equation \ref{eqn:mandel} corresponds to the unlikely case of when the planet is larger (or equal to the same radius) than its host star. 

\begin{table}[bt]
    \caption{{\bf MCMC results.} Only one of the LCOGT data sets gave a reliable solution.  Results of three of the better MObs transits are shown, to demonstrate the lower confidence in the estimated parameter estimates for such data sets (together with an implausibly large `planet'). Uncertainties are $1\sigma$. `Date' is the night of observation. }
    \centering
    \hspace{-2.5cm}
    \begin{tabular}{||l|c|c|c|c|c|l||}
    \hline
Date                & ${r_p}/{r_s}$     & ${r_s}/a$         & $u$               & $\cos{i}$         & $\sigma$                  & Observatory\\
    \hline
04 October 2019     & $0.159 \pm 0.013$ & $0.109 \pm 0.007$ & $0.48 \pm 0.23$   & $0.086 \pm 0.013$ & $0.0036 \pm 0.0001$       & LCOGT  \\
11 October 2020     & $0.35 \pm 0.23$   & $0.14 \pm 0.04$   & $0.55 \pm 0.30$   & $0.16 \pm 0.07$   & $0.010 \pm 0.001$         & MObs \\
20 October 2020     & $0.32 \pm 0.22$   & $0.10 \pm 0.02$   & $0.53 \pm 0.28$   & $0.11 \pm 0.05$   & $0.0058 \pm 0.0005$       & MObs \\
02 January 2021     & $0.33 \pm 0.20$   & $0.11 \pm 0.02$   & $0.58 \pm 0.28$   & $0.11 \pm 0.05$   & $0.0063 \pm 0.0005$       & MObs \\
    \hline 
    \end{tabular} 
    \label{tab:mcmc_results}
\end{table}  

Limb darkening refers to the phenomenon that the brightness of a star appears to decrease from the centre to the edge, or limb, of the observed disk. This occurs because a stellar atmosphere increases in temperature with depth. At the centre of a stellar disk an observer `sees' deeper and hotter layers that emit more light compared to at the limbs, where the upper and cooler layers are seen (which produce less light). The `small planet' approximation was used for the transit model, in that the limb darkening value corresponding to the centre of the planetary disk projected onto the stellar disk was uniformly applied across the stellar area obscured by the planet.  We implemented linear limb darkening for the model to adjust the obscured flux values, i.e., a limb darkening model with only a single term.

Only one of our data sets (LCOGT 04 October 2019) could be reliably fitted with this model, as it had a sufficient signal to noise ratio, a well-defined transit, and sufficient observations before and after the transit so that the out of transit flux levels were well constrained.  Interestingly, we were not able to derive a determinate solution for the 04 October 2019 data set, which by eye appears to be a suitable transit.  This would indicate that we have too many free parameters in the fit, a point we will come back to later in the paper. Table~\ref{tab:mcmc_results} presents results of this fitting and some example MObs fits. Clearly we were asking too much of the MObs data when we included inclination and limb darkening as free parameters, as we have physically unreasonable solutions for these data sets.  {\sc exotic} is a better tool for these high noise data sets.  The HCM fit to the LCOGT data is more reasonable.  

\subsection{Comparison with the Literature}

Hellier {\em et al.} (2016) estimated $r_p / r_s$ as $ 0.166^{+0.059}_{-0.027}$, $\cos{i} = 0.117^{+0.013}_{-0.009}$, and ${r_s}/a = 0.125^{+0.030}_{-0.022}$. These figures are in good agreement with the HMC model fit based on the LCOGT data bar for $\cos{i}$, with the HCM model corresponding to an inclination of $85.07 \pm 0.75$ degrees compared to Hellier {\em et al.'s} value of $83.3^{+0.5}_{-0.8}$ degrees. This is within two standard deviations though. 

A comparison with the results from the {\sc exotic} model for the same data shows that the orbital radius from the HMC model is substantially larger (at $\sim 9.2$ times the stellar radius) as is the planetary radius ({\sc exotic's} $0.131 \pm 0.001 \: r_s$ compared to $0.159 \pm 0.013$). The lack of agreement is puzzling, given that both Hellier {\em et al.} and {\sc exotic} both integrate the limb darkened fluxes obscured by the planetary disk, suggesting that the small planet approximation is not the primary cause of the difference.  

\begin{figure}[!t]  
\begin{subfloat}[TESS Sector 31 Light Curve]{
  \includegraphics[width=0.45\textwidth]{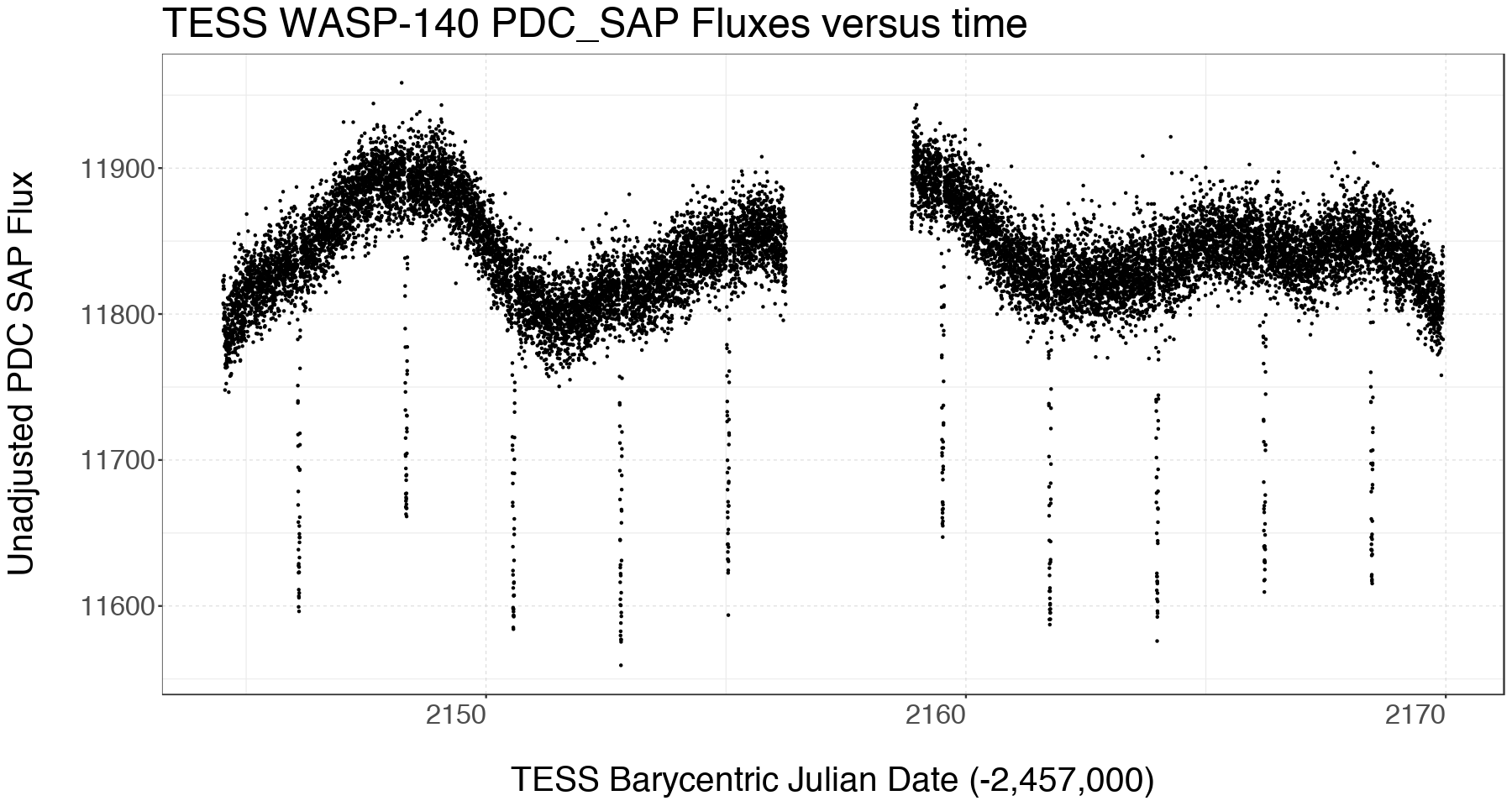}
  \label{fig:tess_1}}
\end{subfloat}\hfil 
\begin{subfloat}[2020-Nov-08 Transit]{
  \includegraphics[width=0.45\linewidth]{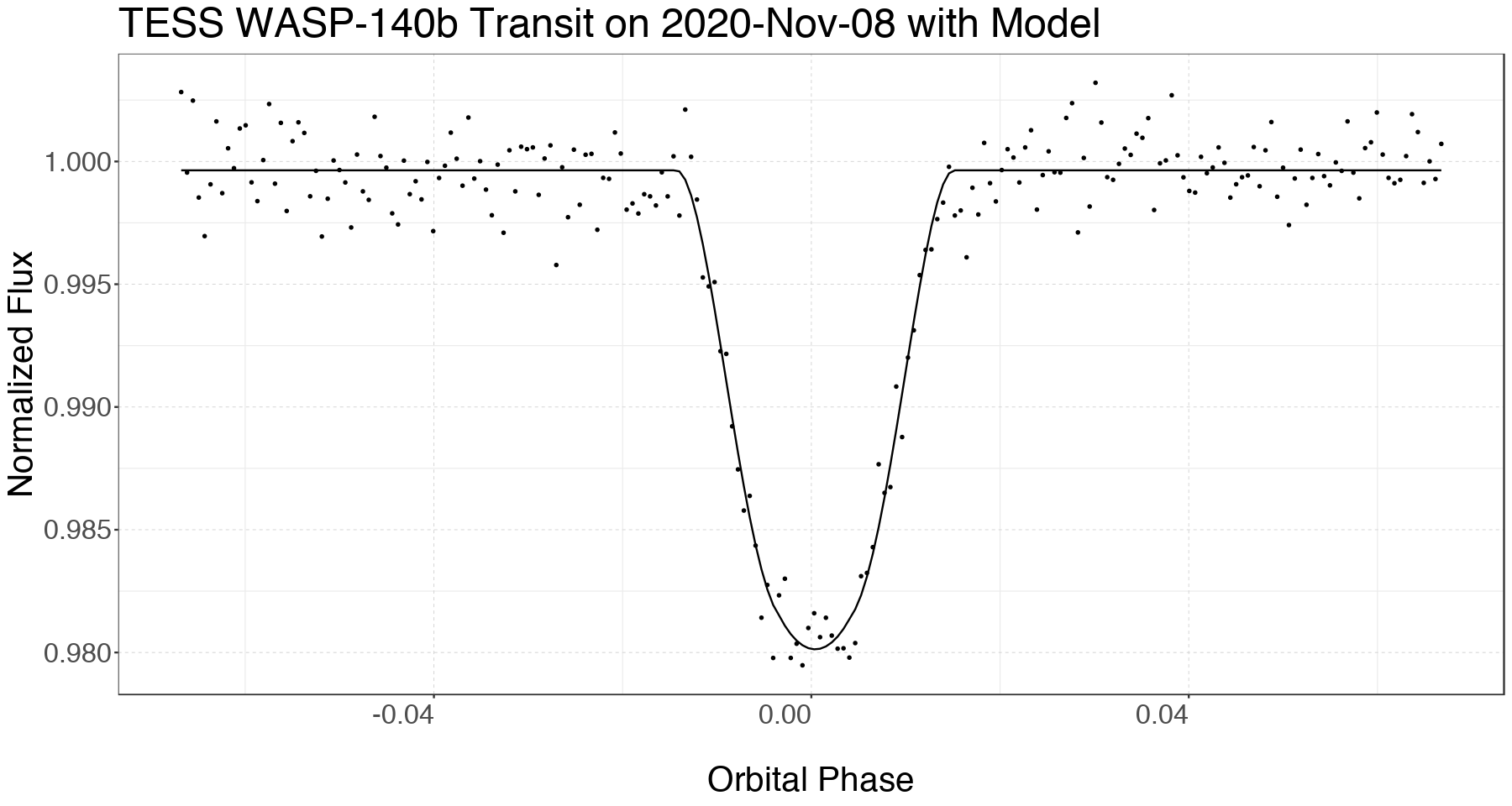}
  \label{fig:tess_2}}
\end{subfloat} 
\caption{The figure on the left (a) shows the non-normalized Pre-search Data Conditioning Simple Aperture Photometry (PDC\_SAP) generated by the TESS team, which has had removed longstanding systematic trends and so provides better data quality than the simple aperture photometry (also available from MAST). Remaining variability is clearly visible, showing these changes are on timescales comparable to that between transits.  Hellier et al. (2016) noted residual variation at a 5-9 milli-magnitude amplitude.  This range is consistent with the observed remaining variability. The figure on the right (b) shows one of these transits plus the optimal model generated by the HMC code. This transit is the second from the left in the data following the break in the middle of Figure \ref{fig:tess_1}.
\label{fig:tess_light_curves}}
\end{figure}

Davoudi et al. (2020) used {\sc EXOFAST} (Eastman {\em et al.}, 2013) to model a clear filter 01 January 2017 transit data set of the system, finding the planet's radius to be $1.1990 \pm 0.0735$ that of Jupiter, which is smaller than Hellier~{\em et al.}'s estimate of $1.44^{+0.42}_{-0.18} \: {\rm R_{J}}$ and this paper's of $1.38^{+0.18}_{-0.17} \: {\rm R_{J}}$ (although within the error ranges). No inclination or orbital radius data were supplied by Davoudi et al., so a comparison is not possible. 

Alexoudi (2022) applied the {\em emcee} Bayesian sampler (Foreman {\em et al.}, 2015) to analyse 28 transits from 3 sectors\footnote{Sector 4 from 18 October 2018 to 15 November 2018, sector 5 from 15 November 2018 to 11 December 2018, and sector 31 from 21 Octo\-ber 2020 to 19 Novem\-ber 2020.} of data collected by the TESS space telescope.  Alexoudi derived an inclination of $84.30 \pm 0.06$ degrees, $r_{s}/a = 0.1166 \pm 0.0008$, and $r_p / r_s = 0.1464 \pm 0.0010$. These values are similar to those of the current paper and Hellier {\em et al.}, but not within formal uncertainties. Alexoudi noted the differences with Hellier {\em et al.}, commenting that these could be due to the higher accuracy of the TESS data.   As a check, we downloaded 2-minute cadence TESS data from MAST (see Figure \ref{fig:tess_1}) and applied the HMC model to a transit (centred on TBJD 2459161.75, see Figure \ref{fig:tess_2}). We found $r_{s}/a = 0.109 \pm 0.008$, $r_p/r_s = 0.163 \pm 0.016$,  and $\cos{i} = 0.089 \pm 0.016$ ($\sim 84.87^{\circ}$). The linear limb darkening coefficient was poorly constrained ($0.48 \pm 0.29$). Our model resulted in a larger planetary radius than Alexoudi's, and very close to those derived from the LCOGT data.

\subsection{Recommendations}

Problems with the other data sets included the lack of sufficient pre-transit data prevented reliable estimates (e.g., the 14 October 2020 data set) while variations in the out-of-transit flux levels prevented a reliable fit to the 28 December 2021 data set. The increased noise of the MObs data compared to LCOGT data also led to less accurate parameter estimates, especially for ratio of the planetary to stellar radii.  It would be interesting to see if additional data processing, such as collection and use of flat fields, would help improve the quality of these data sets.

For transit fittings of this system, we recommend that the pre- and post- transit observations be roughly as long as the actual transit time period, particularly since the host star appears to be active (changing in flux levels) on a short time scale. For instance, the pre-transit flux levels appear to be greater than post-transit for the 28 December 2021 data set, and are a complication for a simple model such as ours.  

A further complication is the use of the small planet approximation for a high inclination orbit such as WASP-140b's; in later projects we intend to apply a graduated limb darkening adjustment to the obscured flux.  There is a clear correlation between $u$ with ${r_p}/{r_s}$ and ${r_s}/a$ (see Figure~\ref{fig:14_oct_19_wasp_140_mcmc}), so locking $u$ to a value based on theory could lead to a tighter confidence interval for these two parameters. The parameter $u$ can also be seen to be poorly defined in Figure~\ref{fig:14_oct_19_wasp_140_mcmc}. This suggests that it could be better to set it to a value using theory and include $u$ as a fixed (rather than a free) parameter. See Banks \& Budding (1990) for further discussion of the information content of data and the question of over-parameterization. Finally, WASP-140b transits close to the stellar limb where the gradient will be strongest in the limb darkening, further supporting the conclusion above.

The signal to noise ratio is clearly important for transit fitting, affecting the accuracy of the MObs fits by our model. Observations with the LCOGT (similar to those presented here) appear to have sufficient ``information content'' to support the HMC model, providing sufficient data about the shoulders of the eclipse are collected for accurate estimation of the out-of-transit flux level.

\section{Summary}

This paper presented MCMC modeling of transits of WASP-140b, collected using robotic telescopes of the MObs and LCOGT. These data included a transit in December 2021 collected by the authors. We coded a fitting function based on the equations of Mandel~\& Agol (2002) and coupled this with Bayesian optimization. Together with the {\sc exotic} analysis program, two MCMC-based optimization models have been applied to these transits, deriving estimates for the times of mid-transit as well as physical parameters of the system.  Inspection of the mid-transit times revealed a linear period with no statistical evidence from the data of transit time variations, i.e., no evidence for the gravitational influence of a non-transit planet on the orbit of WASP-140b. 

Results from the two analysis programs ({\sc exotic} and HMC) were in good agreement, indicating that the radius for WASP-140b to be $1.38^{+0.18}_{-0.17}$ Jupiter radii, with the planet orbiting its host star in $2.235987 \pm 0.000008$ days at an inclination of $85.75 \pm 0.75$ degrees. The derived parameters are in formal agreement with the discovery paper of Hellier {\em et al.} (2016), and somewhat larger than a recent independent study based on photometry by the TESS space telescope (Alexoudi, 2022). 

We were probably too ambitious in our selection  of an exoplanet with a high inclination orbit about a host star itself with rapidly changing flux levels (to apply a high parameter model such as the HMC model), but that is part of the learning process. Application of techniques such as Gaussian Processes to model out the host star variations would be a good next step, which would allow the combination together of multiple transits which could be binned together to increase the signal to noise ratio and strengthen the information content of the data. We also plan to use our HMC model on more simple systems, such as Kepler 1\footnote{See, e.g., Ng et al. (2021) who applied the Mandel~\& Agol (2002) models, MCMC, and Gaussian Processes to Kepler space telescope data of Kepler-1b and other systems.} which do not have such active host stars and orbits with inclinations closer to 90 degrees, where the model's deficiencies will be less and the correlation between limb darkening and inclination less confounding. Having made these comments, we still recommend that programming a simple model such as Mandel \& Agol (2002) and coupling this with an optimizer is a useful learning exercise, and makes for a useful student project.  Our points are rather to choose a more quiet system than the one we did, and to either implement improved handling of limb darkening for highly tilted systems or to choose an exoplanet with an orbit closer to $90^{\circ}$ inclination as well as being somewhat smaller relative to its host star (so that the small planet approximation is more valid). If investigation of TTVs is the primary goal of the project, then {\sc exotic} is an excellent tool for such work.

\newpage
\begin{acknowledgments}

This publication makes use of the EXOTIC data reduction package from Exoplanet Watch, a citizen science project managed by NASA’s Jet Propulsion Laboratory (JPL) on behalf of NASA’s Universe of Learning and which is supported by NASA under award number NNX16AC65A to the Space Telescope Science Institute. We are grateful for observing time on the Las Cumbres Observatory Global Telescope (LCOGT) Network, and to Rachel Zimmerman Brachman (JPL) for making available this opportunity. We thank the LCOGT for making available archival data. We also thank the  Harvard-Smithsonian Institute for Astrophysics for the MicroObservatory data kindly made available by Frank Sienkiewicz.  This research has made use of the NASA Exoplanet Archive, which is operated by the California Institute of Technology, under contract with the National Aeronautics and Space Administration under the Exoplanet Exploration Program. We thank the University of Queensland for collaboration software. This paper includes data collected by the TESS mission and obtained from the MAST data archive at the Space Telescope Science Institute (STScI). STScI is operated by the Association of Universities for Research in Astronomy, Inc., under NASA contract NAS 5–26555. We thank the anonymous referee for their comments and guidance which improved the paper.

\end{acknowledgments}


\allauthors
\end{document}